\documentclass[a4paper,10pt,twoside]{cpc-hepnp}

\usepackage{multicol}
\usepackage{graphicx}
\usepackage{booktabs}
\usepackage{amssymb,bm,mathrsfs,bbm,amscd}
\usepackage[perpage,symbol]{footmisc}

\usepackage[tbtags]{amsmath}
\usepackage{lastpage}
\usepackage{lineno}
\usepackage{fancyhdr}

\begin{document}

\fancyhead[c]{\small Submitted to Chinese Physics C}

\title{Study of the calibration of X-T relation for the BES\uppercase\expandafter{\romannumeral3} drift chamber \thanks{Supported by National Natural Science Foundation of China (11205184,11205182,11121092,11179020) and Joint Funds of National Natural Science Foundation of China (U1232201)}}

\author{
      KANG Xiao-Lin$^{1,2;1)}$\email{kangxl@mail.ihep.ac.cn}
\quad WU Ling-Hui$^{1;2)}$\email{wulh@mail.ihep.ac.cn}
\quad WU Zhi$^{1}$
\quad LUO Tao$^{3}$\\
\quad HU Chen$^{4}$
\quad WANG Hai-Xia$^{5}$
\quad FANG Shuang-Shi$^{1}$
\quad HE Kang-Lin$^{1}$\\
\quad LI Wei-Dong$^{1}$
\quad LI Wei-Guo$^{1}$
\quad MAO Ze-Pu$^{1}$
\quad WANG Liang-Liang$^{1}$\\
\quad YUAN Ye$^{1}$
\quad ZHANG Yao$^{1}$
}
\maketitle

\address{
$^1$ Institute of High Energy Physics, Chinese Academy of Sciences, Beijing 100049, China\\
$^2$ University of Chinese Academy of Sciences, Beijing 100049, China\\
$^3$ University of Hawaii, Honolulu, Hawaii 96822, USA\\
$^4$ Nanjing Normal University, Nanjing 210023, China \\
$^5$ Central China Normal University, Wuhan 430079, China\\
}

\begin{abstract}
This paper introduces the calibration of the time-to-distance relation for the BES\uppercase\expandafter{\romannumeral3} drift chamber. The parameterization of the time-to-distance relation is presented. The studies of left-right asymmetry and the variation with the entrance angle are performed. The impact of dead channels on the time-to-distance relation is given special attention in order to reduce the shifts of the measured momenta for the tracks passing near dead cells. Finally we present the spatial resolution (123 $\mu$m) for barrel Bhabha events ($|\cos\theta|<0.8$) from J/$\psi$ data taken in 2012.
\end{abstract}



\begin{multicols}{2}

\section{Introduction}
The Beijing Spectrometer \uppercase\expandafter{\romannumeral3} (BES\uppercase\expandafter{\romannumeral3}) \cite{1}, which operates at the upgraded Beijing Electron-Positron Collider (BEPC\uppercase\expandafter{\romannumeral2}) \cite{2}, aims at precision measurements and new physics searches in $\tau$-charm energy region. The central tracking system of BES\uppercase\expandafter{\romannumeral3} is a cylindrical multilayered drift chamber
with 24 stereo layers and 19 axial layers in total.
A helium based gas mixture of  He/C$_3$H$_8$ (60/40) was chosen to minimize the effect of multiple scattering. The BES\uppercase\expandafter{\romannumeral3} drift chamber operates in a 1 T magnetic field and is required to provide good spatial resolution (130 $\mu$m) and good momentum resolution (0.5\%@1 GeV/c).

In order to meet these requirements, precise offline tracking calibration and alignment are essential. The offline calibration software for the drift chamber, which is developed in the BES\uppercase\expandafter{\romannumeral3} Offline Software System (BOSS) \cite{3}, contains the calibration of the time-to-distance relation (X-T relation), time offset (T0) and time walk effects (Q-T relation). Time offset, which is determined cell by cell, is corrected by the difference between the measured and estimated drift distance \cite{4,5}. The calibration of time walk effects is determined from the time shift as a function of ADC output (Q) for each layer \cite{4}. The calibration of the X-T relations is more complicated than those for time offset and time walk effects.
We use different X-T relations for different layers to account for the variations in cell size and electric field. In order to achieve good spatial resolution, we also take into account the left-right asymmetry, the dependence on the incident direction of the track and other factors such as dead channels.

In this paper we introduce how the X-T relation is parameterized, describe studies of the above-mentioned issues that affect X-T relations and give the present spatial resolution of the drift chamber for a J/$\psi$ data set that was collected in 2012. The Bhabha events we used are detected in the barrel region of the detector ($|\cos\theta|<0.8$).

\section{Cell geometry and the parameterization of the X-T relation}

The drift cell of the BES\uppercase\expandafter{\romannumeral3} drift chamber has an almost square trapezoidal shape. A total of 6796 cells are arranged in 43 cylindrical layers. Each sense wire is surrounded by eight field wires. The cell size is approximately 12$\times$12 mm$^2$ for the inner eight layers and 16.2$\times$16.2 mm$^2$ for the others. The sense wires are connected with high voltage cables via an insulating feedthrough and all of the field wires are kept at ground.

The X-T relation is affected by the gas mixture, and the electrostatic and magnetic fields. A typical scatter plot of the drift distance as a function of the drift time obtained using fitted tracks from BES\uppercase\expandafter{\romannumeral3} events is shown in Fig.~\ref{xtintr}. The X-T relation is smooth over most of the cell's extent. In the region near the cell edge, winding drift lines result in long drift times. The parameterization of the X-T relation is the first step in the calibration procedure. In the Belle experiment, the X-T relation of the Central Drift Chamber (CDC) is fitted with up to fifth-order polynomials \cite{6}; the KLOE drift chamber uses up to fifth-order Chebychev polynomials \cite{7}.

\begin{center}
  \includegraphics[width=6cm,height=4cm]{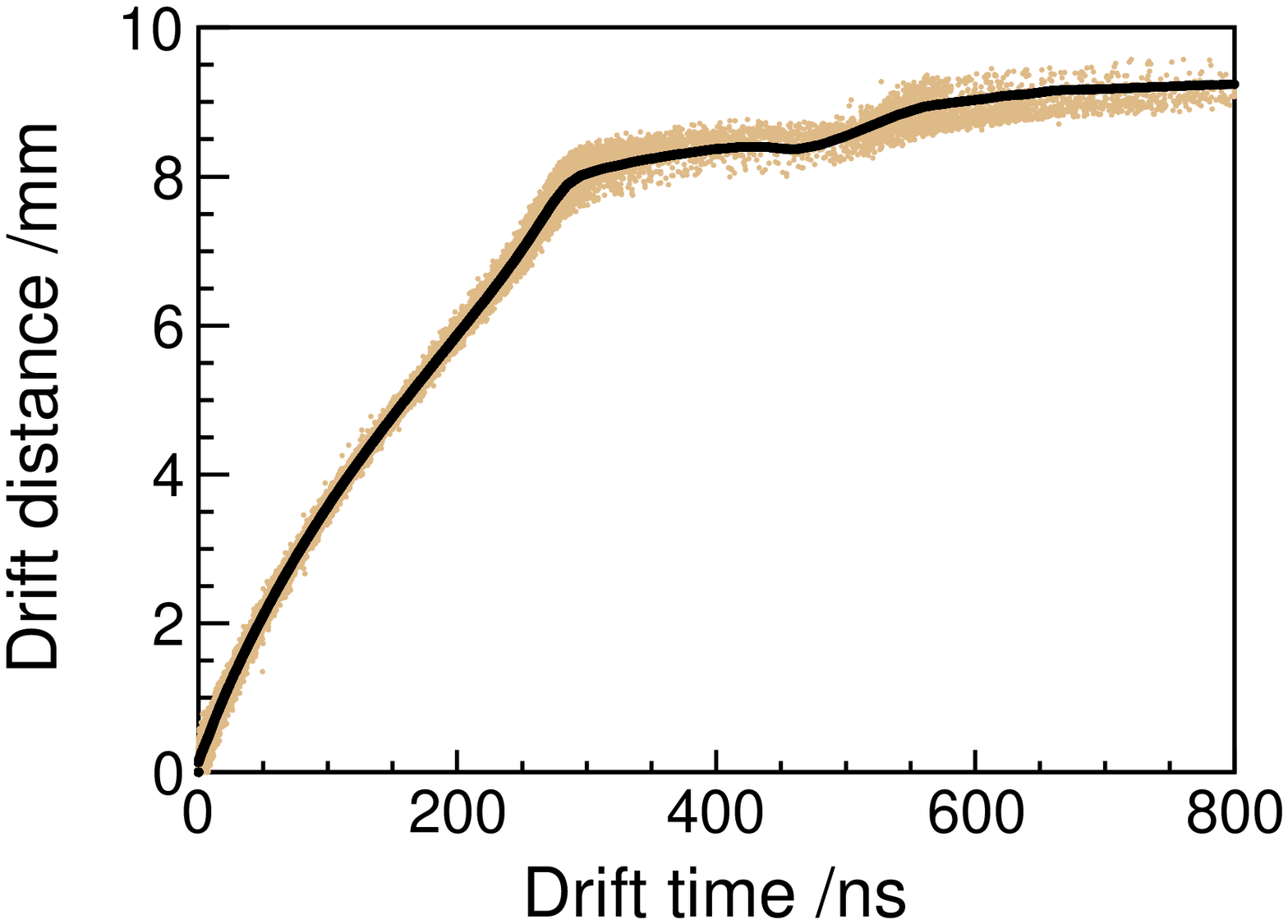}
  \figcaption{\label{xtintr} A typical X-T relation for BES\uppercase\expandafter{\romannumeral3} drift chamber. The scattered points are predicted from track fitting and the black line is obtained from calibration.}
\end{center}

Since the BES\uppercase\expandafter{\romannumeral3} chamber has a similar cell geometry to that of the Belle CDC, we tried to use the same fitting function. However, it does not fit the region near the cell edge very well. Instead we use a numerical method rather than an analytical expression, similar to what the CLEO \uppercase\expandafter{\romannumeral3} drift chamber group did \cite{8}. A time to distance table is saved in a calibration file and an interpolation procedure is used in the reconstruction. In Fig.~\ref{xtintr}, the scattered points are obtained from track fitting while the black line is the result of calibration, where there is good agreement. Figure \ref{resiintr} presents the mean value of residual distribution as a function of drift time. The shift shown in the figure is very small compared with the spatial resolution, which can be neglected.

\begin{center}
   \includegraphics[width=6cm,height=4cm]{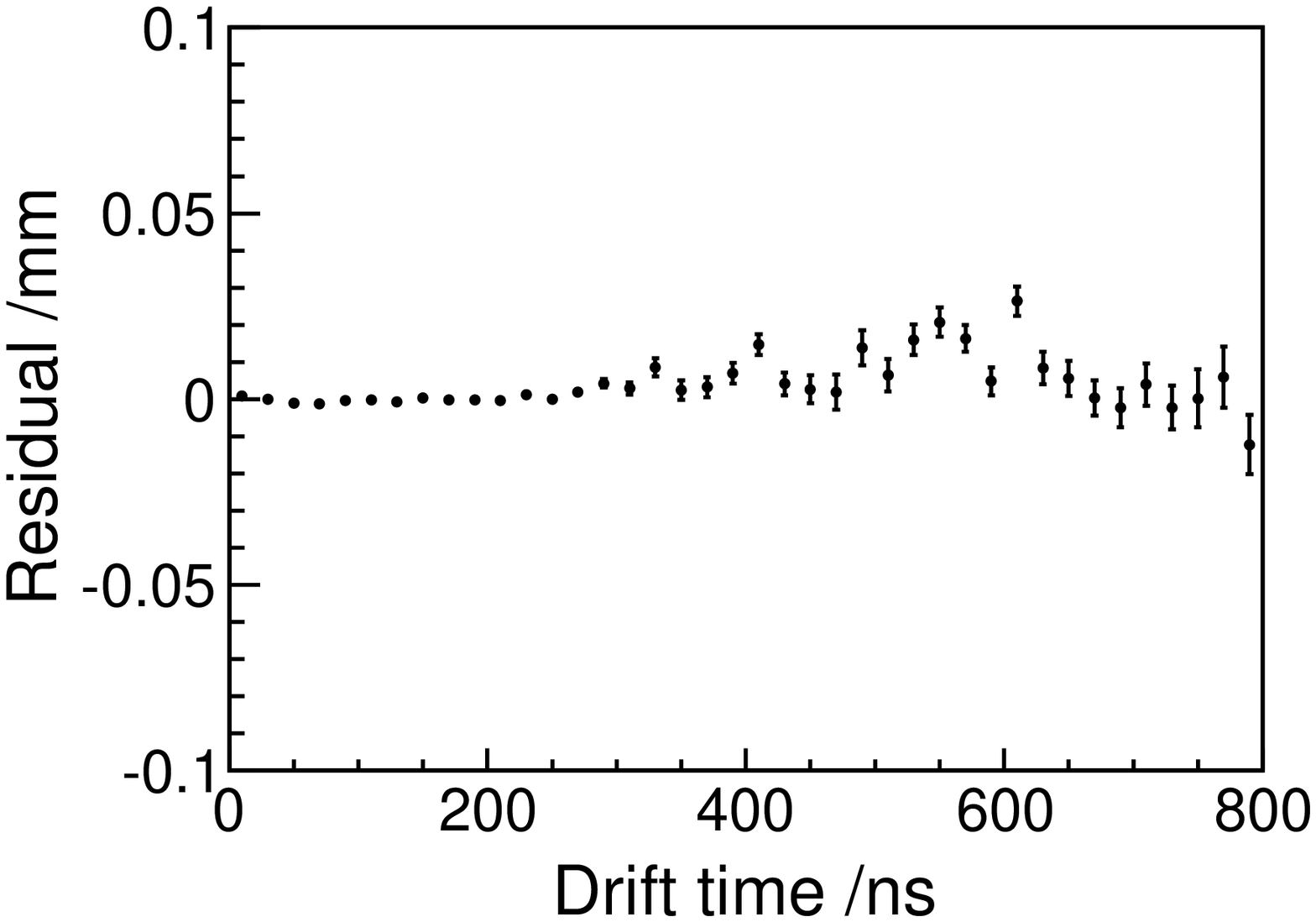}
  \figcaption{\label{resiintr} The mean value of residual distribution versus drift time.}
\end{center}

\section{Left-right asymmetry}
The trapezoidal shape of the cell in the cylindrical drift chamber can cause a radial electric field asymmetry. For cells in non-boundary layers, the asymmetry is small and can be further reduced by optimizing the space between the sense and field layer. But for cells in boundary layers (including axial-stereo boundaries), a large asymmetry in the radial electric field can lead to a big difference between the left- and right-side X-T relations \cite{9}. Although a compensating 100$\sim$300 V potential on the boundary field wires would effectively minimize this asymmetry \cite{10}, it was not applied in order to avoid the reduction of anti-noise capabilities. As a result, we calibrated the X-T relations for left and right separately. Figure \ref{lrasy} shows the difference between the left and right drift distance as a function of drift time for cells in boundary and non-boundary layers. The left-right asymmetry in cells from boundary layers can be as high as 0.4 mm.

\begin{center}
  \includegraphics[width=6cm,height=4cm]{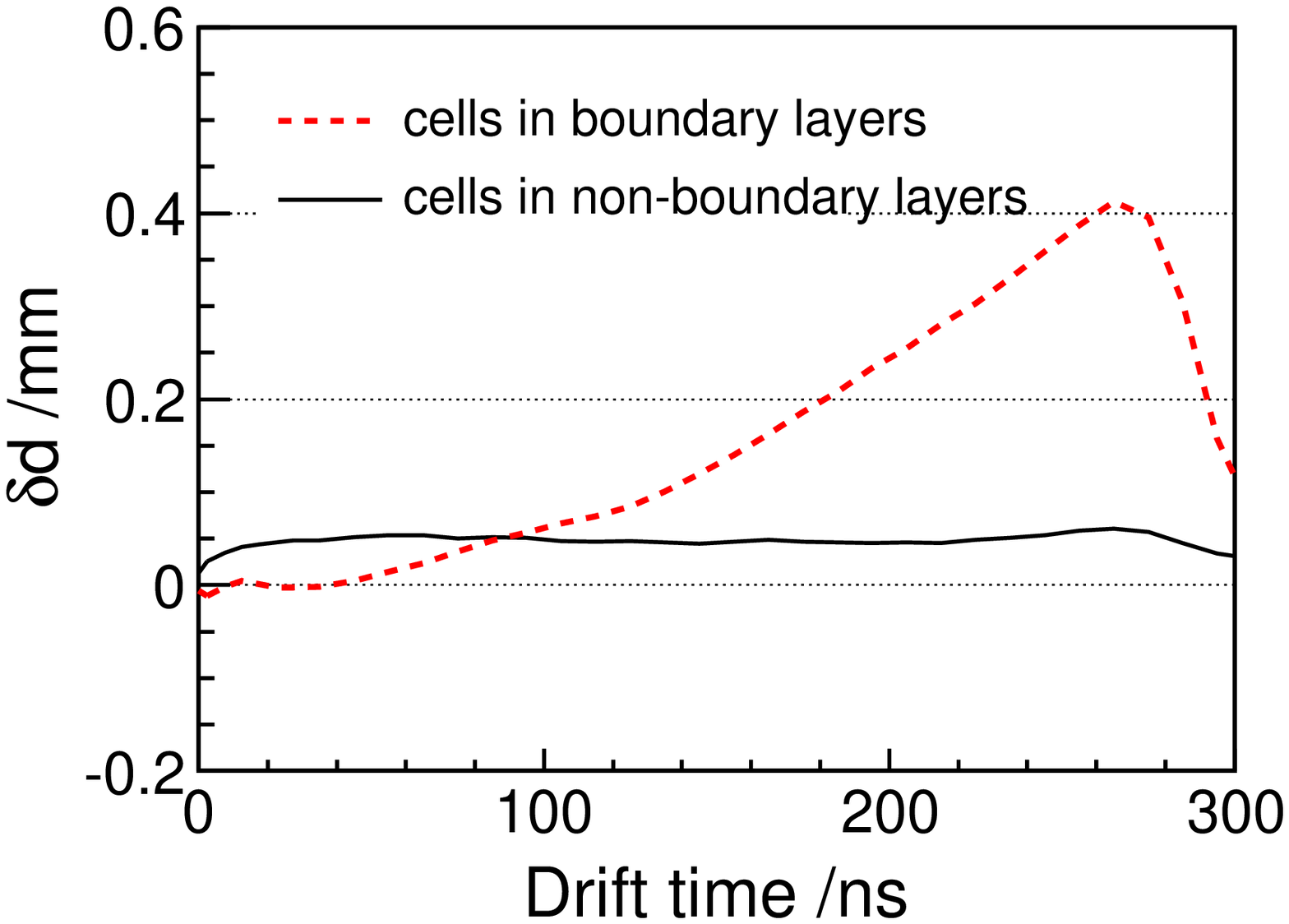}
  \figcaption{\label{lrasy} Left-right asymmetry of the X-T relations ($\delta$d means the difference between the left and right drift distance).}
\end{center}

\section{Variation of the X-T relation with large entrance angles}
In trapezoidal-shaped drift cells, the X-T relations depend on the incident direction of the track, i.e. entrance angle $\alpha$, which is defined as the difference of the azimuth angle between the track (the tangent direction of the hit) and a vector between the origin and the wire in the r-$\phi$ plane, i.e. $\phi_{track}$-$\phi_{wire}$. The full domain of $\alpha$, which depends upon the transverse momentum of the track, is [-$\pi$/2,$\pi$/2]. Figure \ref{entrangle} shows the predicted $\alpha$ in each layer based on fitted ideal helices for different momenta tracks that pass through the origin. For a given track, $|\alpha|$ increases with the layer radius. Within the same layer, $|\alpha|$ varies inversely with the transverse momentum.

\begin{center}
  \includegraphics[width=6cm, height=4cm]{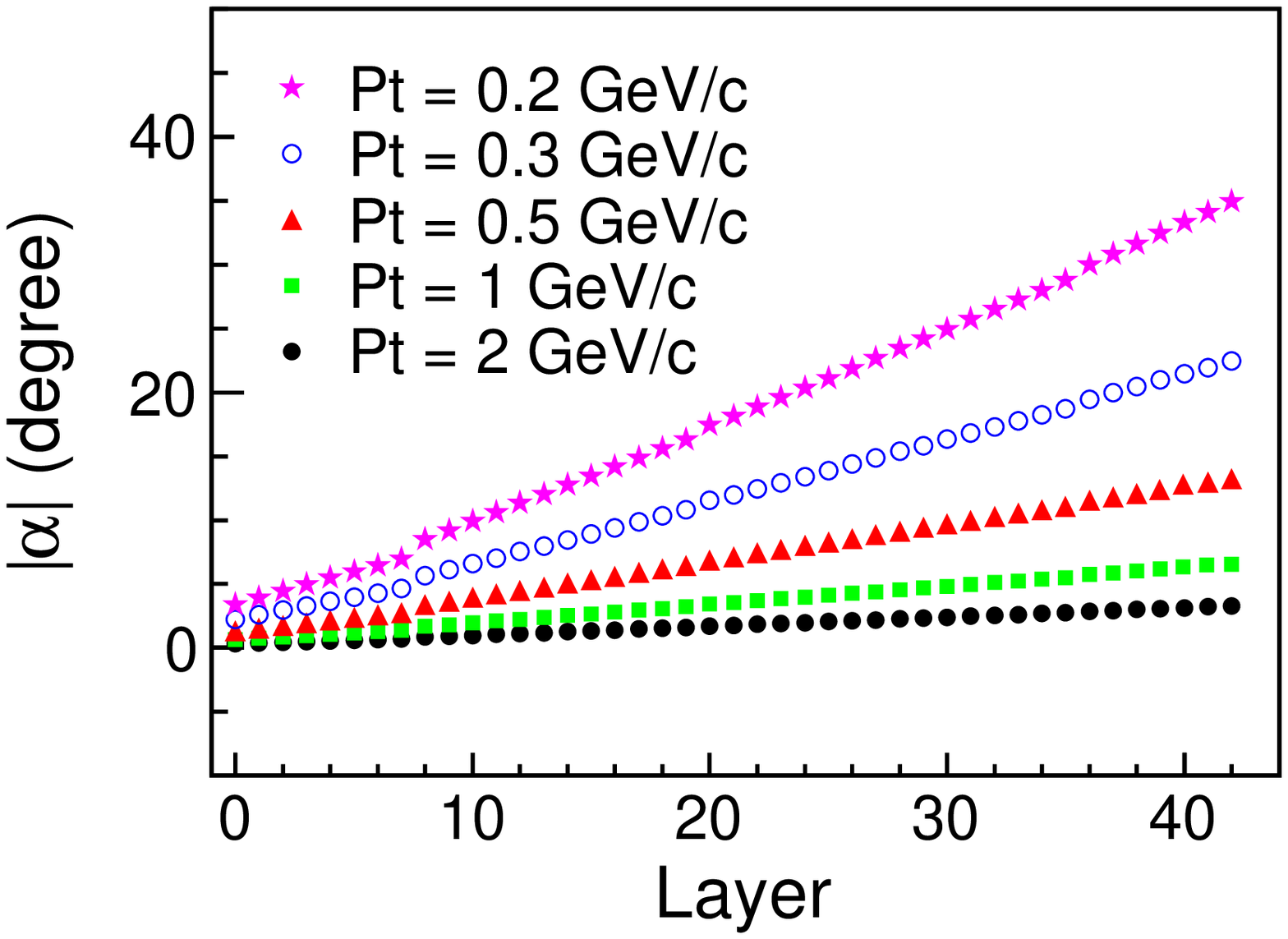}
  \figcaption{\label{entrangle} Entrance angles in different layers for different momentum tracks.}
\end{center}

Bhabha and Dimuon samples, which consist of high momentum tracks, are usually used in calibration. If we use X-T relations calibrated from those samples in physical events which contain lots of low momentum tracks, a bias is unavoidable. The bias can be found in residual distributions, as shown in Fig.~\ref{resipi} (a). These residual shifts can be as high as 0.15 mm for distances far from sense wire.

\begin{center}
  \includegraphics[width=6cm,height=4cm]{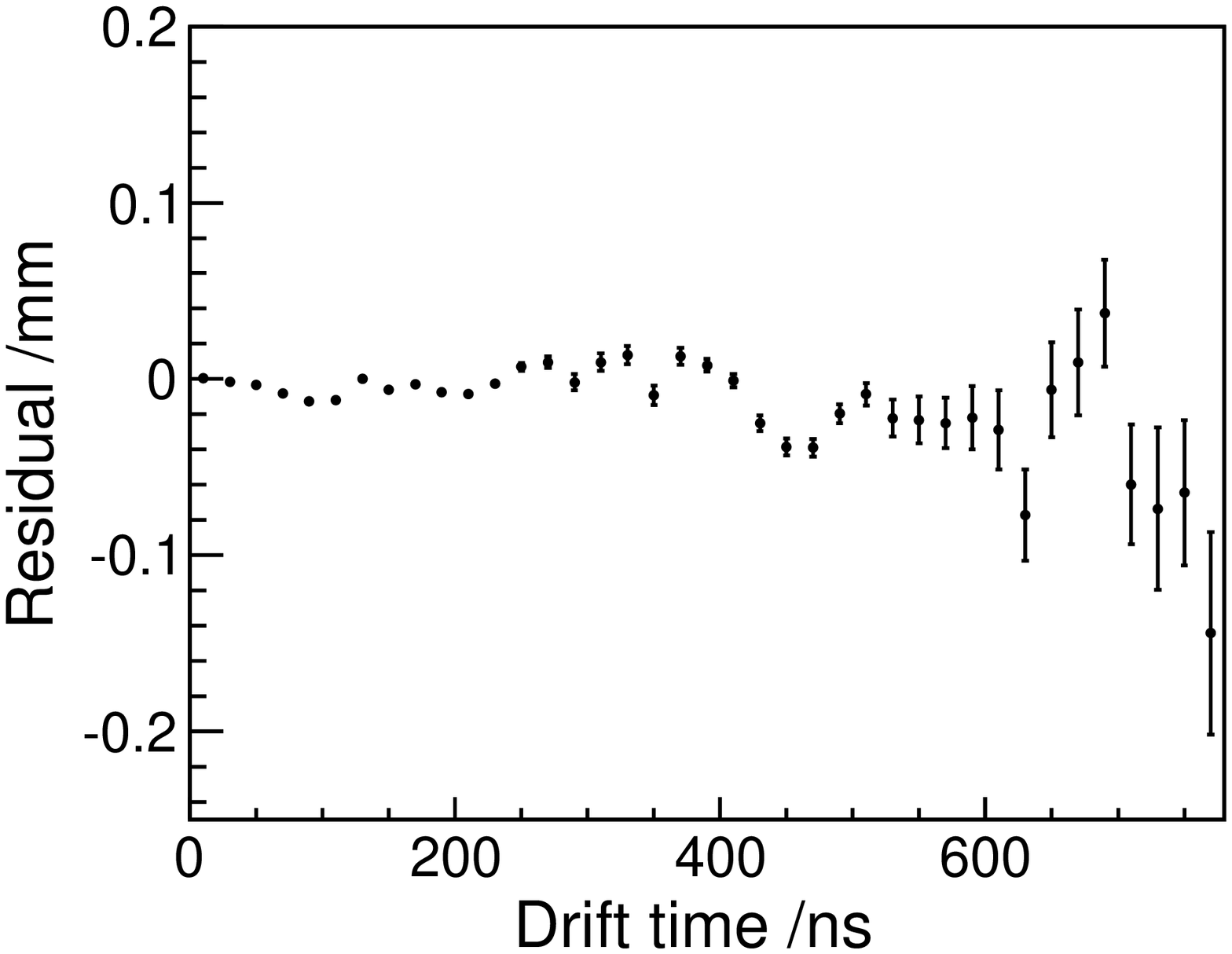}\put(-130,90){(a)}

  \includegraphics[width=6cm,height=4cm]{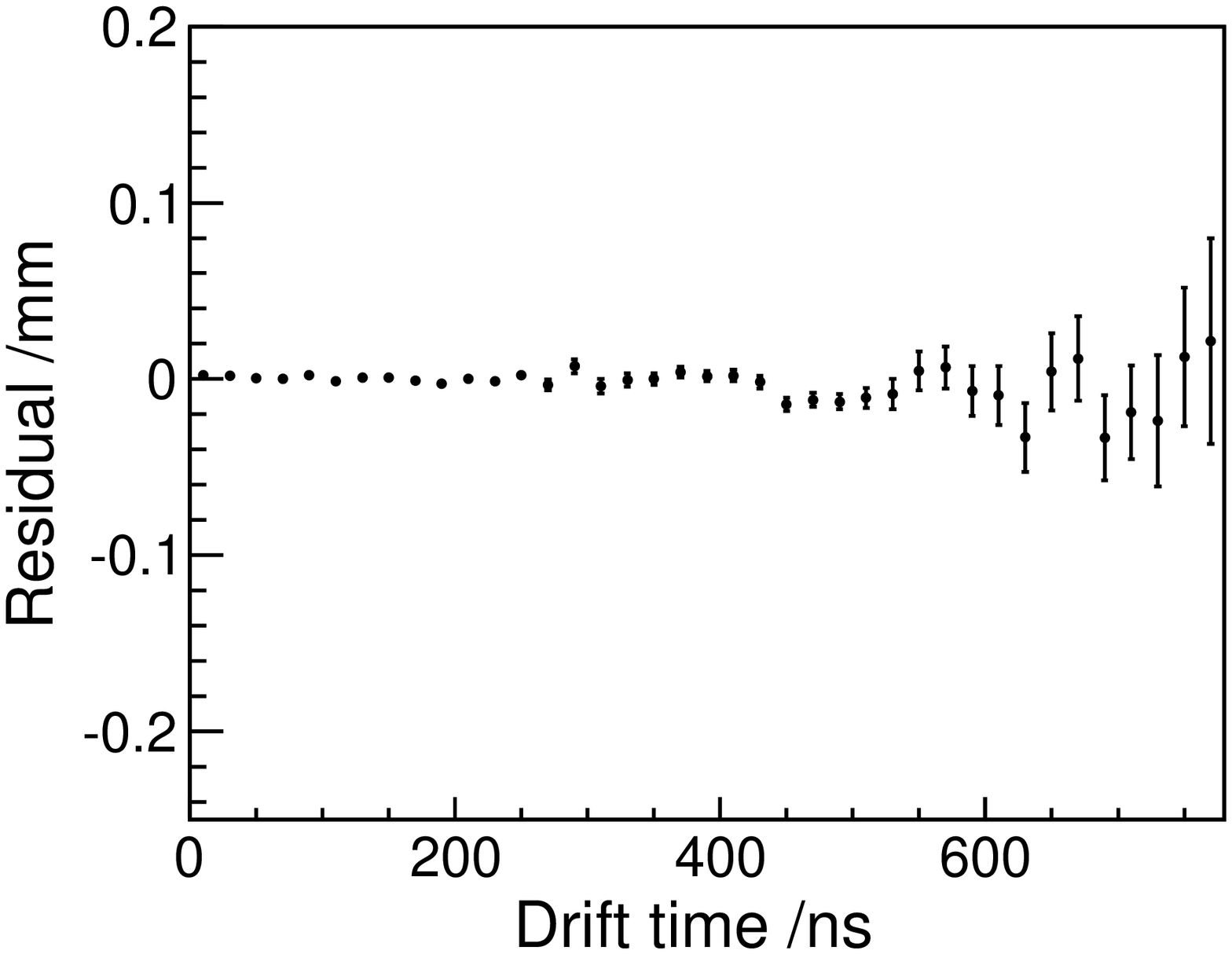}\put(-130,90){(b)}
  \figcaption{\label{resipi} The mean value of residual distribution as a function of drift time with $P_t$ less than 0.5 GeV/c. (a) Using X-T relations calibrated from Bhabha samples. (b) Using X-T relations calibrated from $\pi$ and Bhabha samples. }
\end{center}

This makes it necessary to incorporate both low momentum and Bhabha track samples in the calibration procedure. For example, for the J/$\psi$ data set, we choose $\pi$ tracks from J/$\psi$$\rightarrow$$\rho$$\pi$, for which the pion purity is 99.9\%. Separate X-T relations are determined for 10 degree steps in $\alpha$. Figure \ref{resipi} (b) shows the mean value of the residual distribution as a function of drift time after calibration, in which the bias become much smaller. Figure \ref{xtentr} (a) shows the X-T relations for different entrance angles in layer 30. Figure \ref{xtentr} (b) shows the differences in X-T relations between two sets of adjacent $\alpha$ steps. It is evident that the differences are very small near the sense wires, but grow with the increasing drift times up to as much as several hundred microns.

\begin{center}
  \includegraphics[width=6cm,height=4cm]{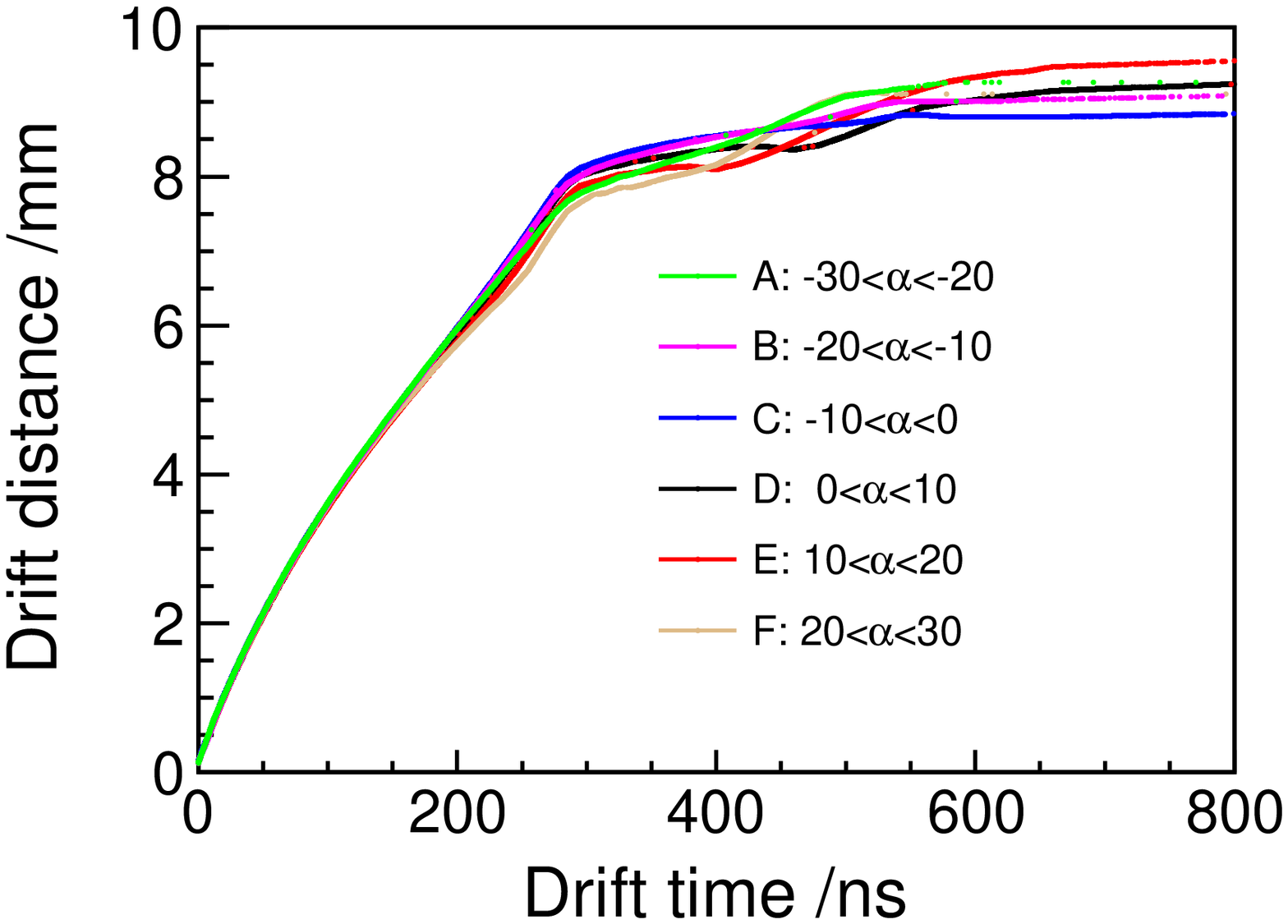}\put(-130,90){(a)}

  \includegraphics[width=6cm,height=4cm]{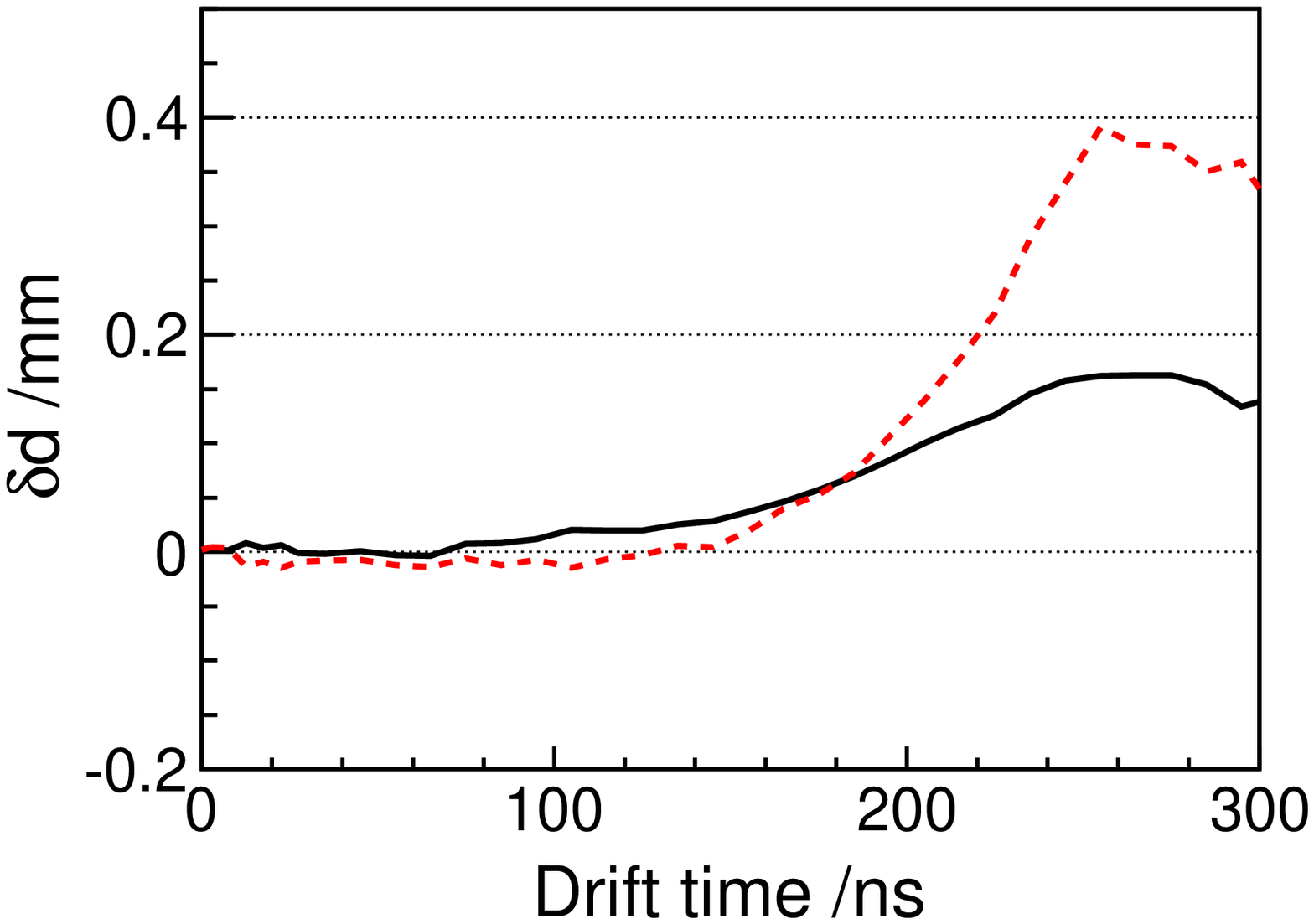}\put(-130,90){(b)}
  \figcaption{\label{xtentr} (a) X-T relations for different entrance angles in layer 30. (b) The difference of drift distance between E and D (solid), F and D (dashed), as a function of drift time.}
\end{center}

\section{Impact of dead channels on X-T relation}

Unavoidably, the BES\uppercase\expandafter{\romannumeral3} drift chamber has some dead channels that develop during data taking periods, usually due to some bad preamplifier cards. Each preamplifier card covers 8 channels. For example, during J/$\psi$ data set collected in 2012, a bad preamplifier caused 8 dead channels (cells 161 through 168 in layer 43). The sense wires for these dead channels were disconnected from their high voltage cables.

The loss of high voltage on a sense wire has an impact on the electrostatic field in surrounding cells that will affect the drift behavior of ionized electrons and result in the variation of drift lines and X-T relations. (In this case cells 160 through 169 in layer 42 were affected.) The different charge distribution for normal cells and the cells affected by dead channels, shown in Fig.~\ref{adc}, reflects the variation of the electric field and gas gain.

\begin{center}
  \includegraphics[width=6cm,height=4cm]{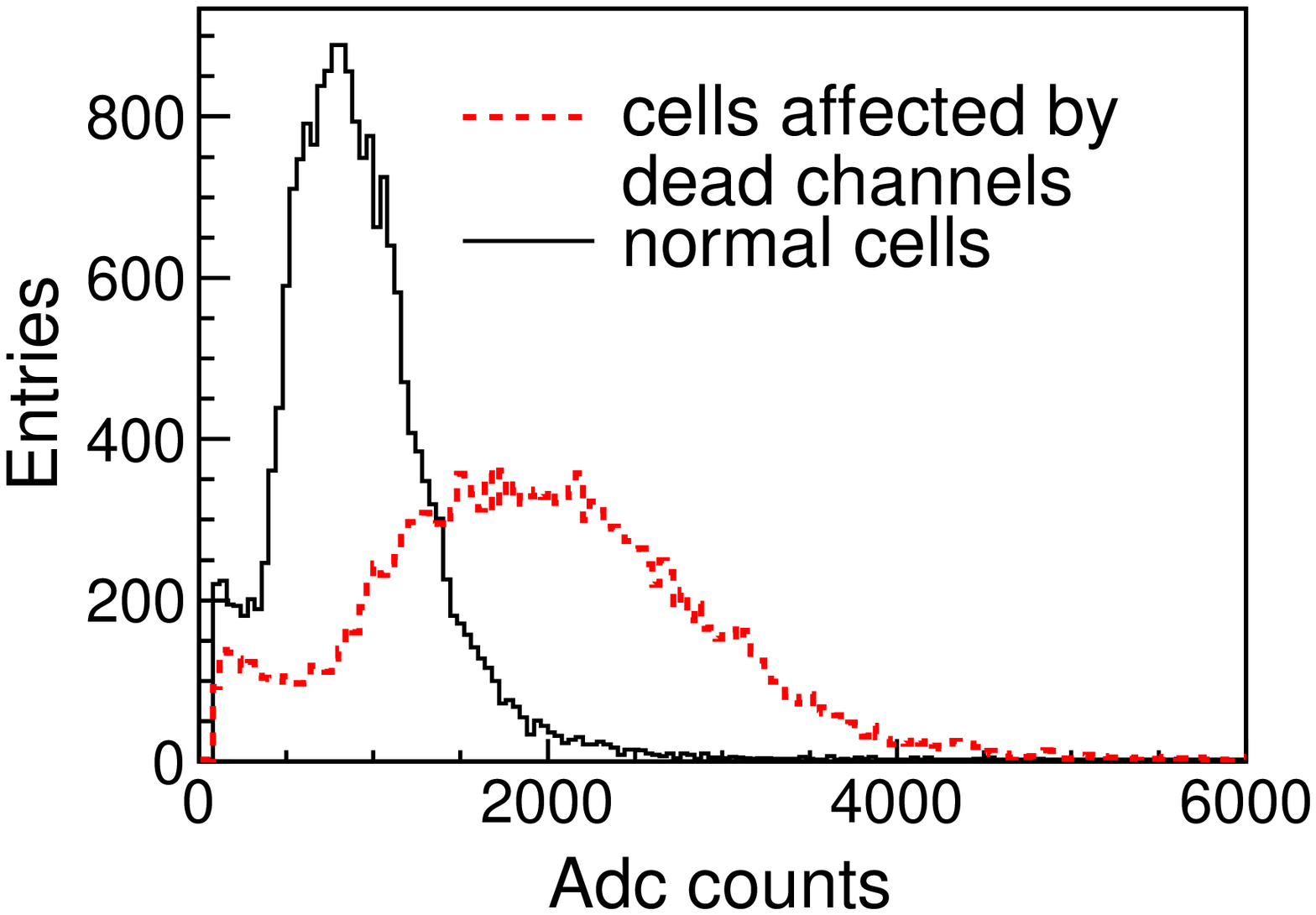}
  \figcaption{\label{adc} Charge distributions.}
\end{center}

The impact on the X-T relation was studied by a Garfield \cite{11} simulation. The difference of drift distance between normal cells and the cells affected by dead channels as a function of drift time is shown in Fig.~\ref{garsim}. The difference is very small near the sense wire, but increases rapidly with drift time.

\begin{center}
  \includegraphics[width=6cm,height=4cm]{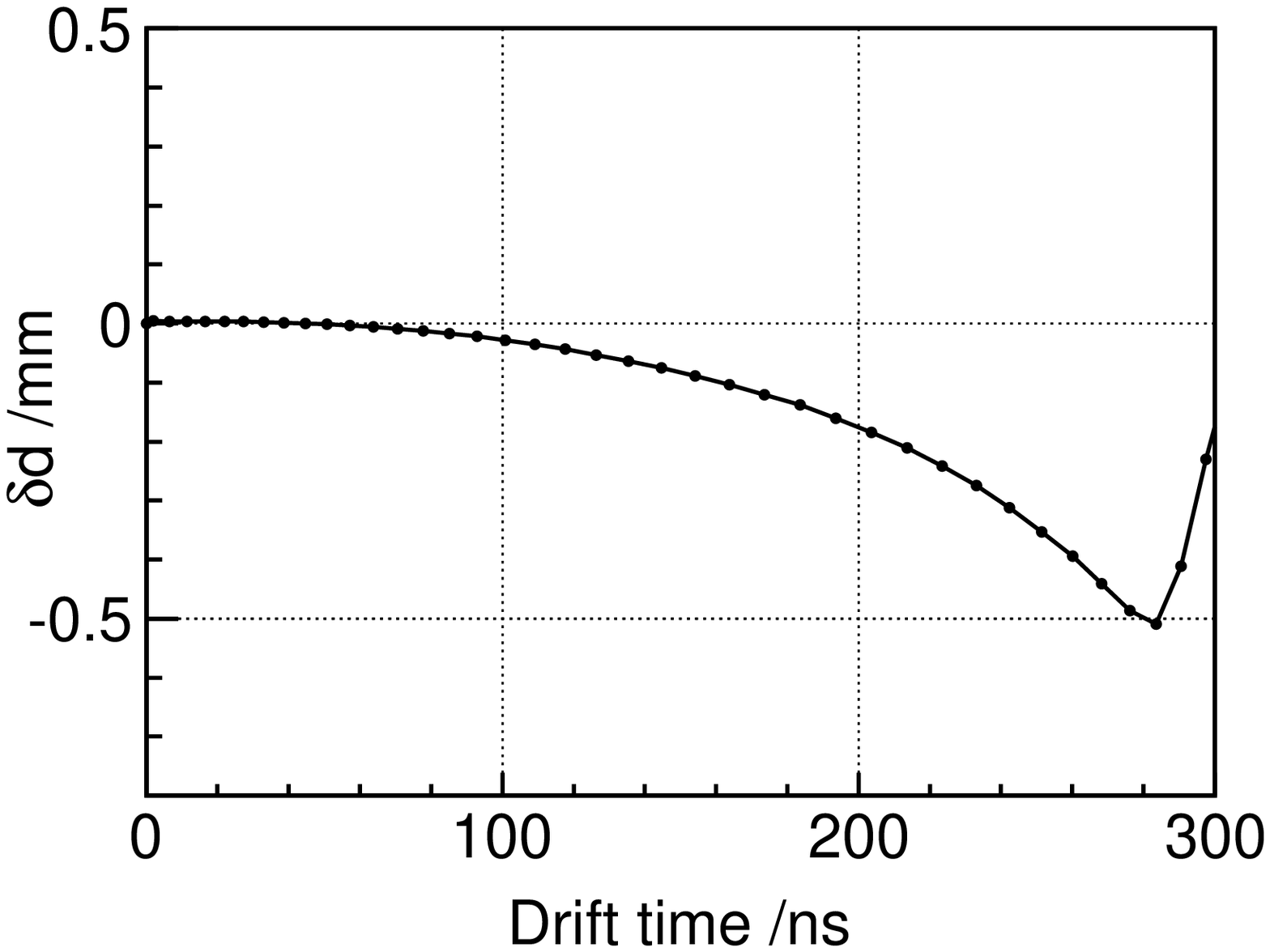}
  \figcaption{\label{garsim} The difference of the X-T relations between normal cells and the cells affected by dead channels simulated by Garfield.}
\end{center}

If we ignore the dead channels in the calibration, an incorrect X-T relation will be obtained for those cells affected by dead channels. This will cause bias on track fitting, which can produce an offset in the residual distribution, as shown in Fig.~\ref{resi_deadch} (a). The bias can cause a big difference in the reconstructed momentum for electrons and positrons that pass through the affected region ($\phi\simeq 2.75$ rads), as shown in Fig.~\ref{resi_mom} (a).

\begin{center}
  \includegraphics[width=6cm,height=4cm]{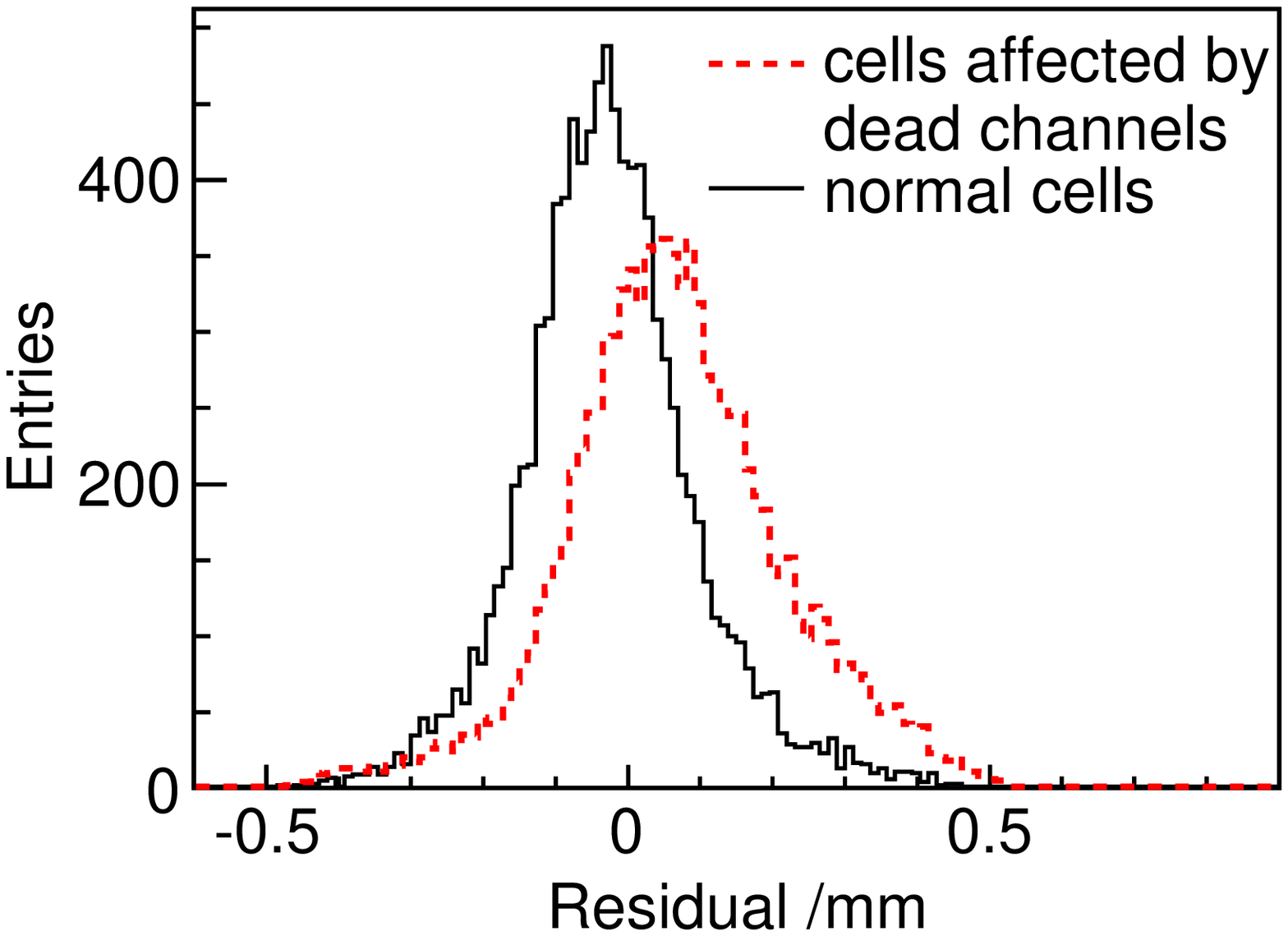}\put(-130,90){(a)}

  \includegraphics[width=6cm,height=4cm]{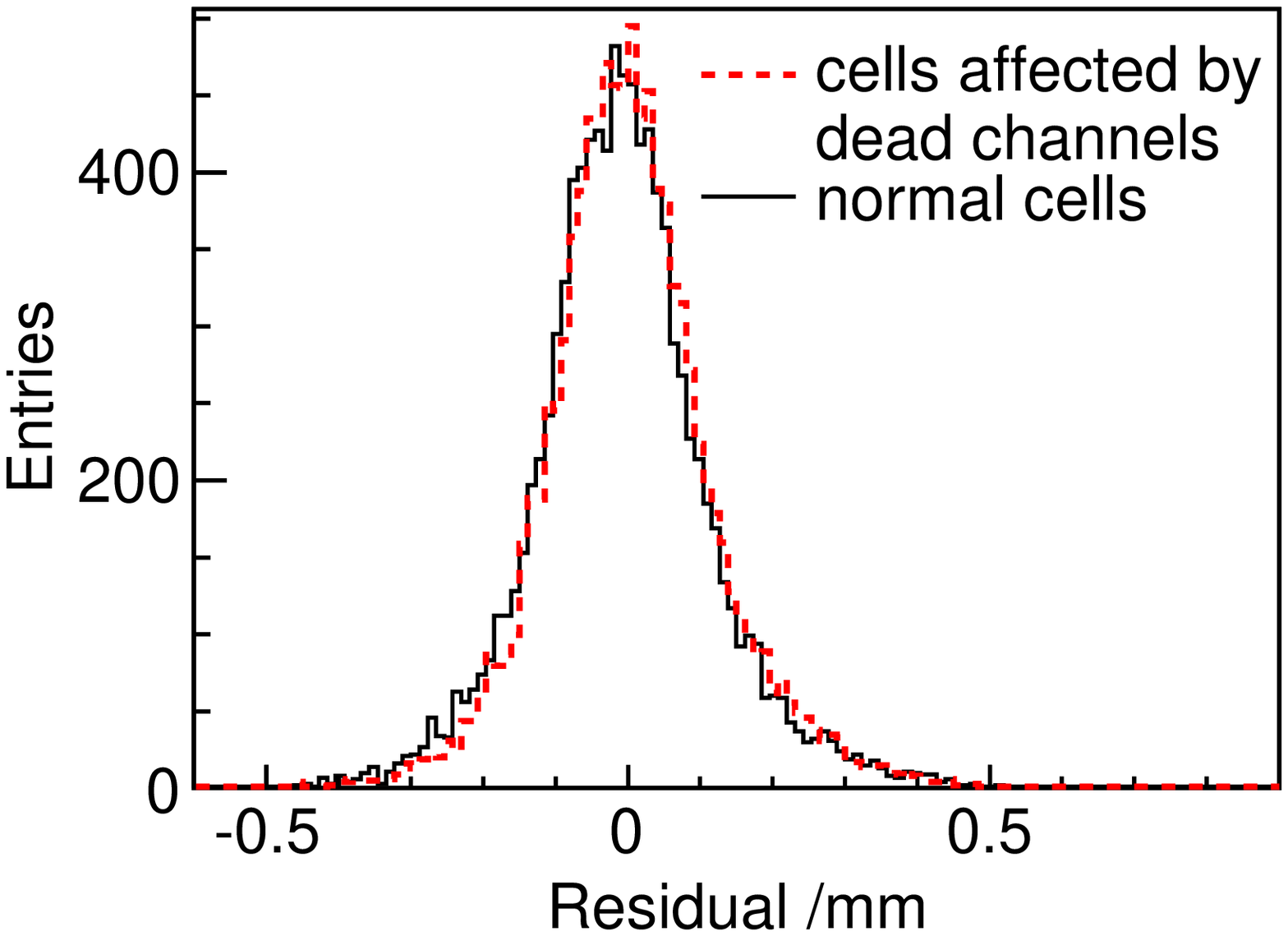}\put(-130,90){(b)}
  \figcaption{\label{resi_deadch} Residual distribution for normal cells (solid line) and cells affected by dead channels (dashed line), (a) before recalibration and (b) after recalibration.}
\end{center}

\begin{center}
  \includegraphics[width=6cm,height=4cm]{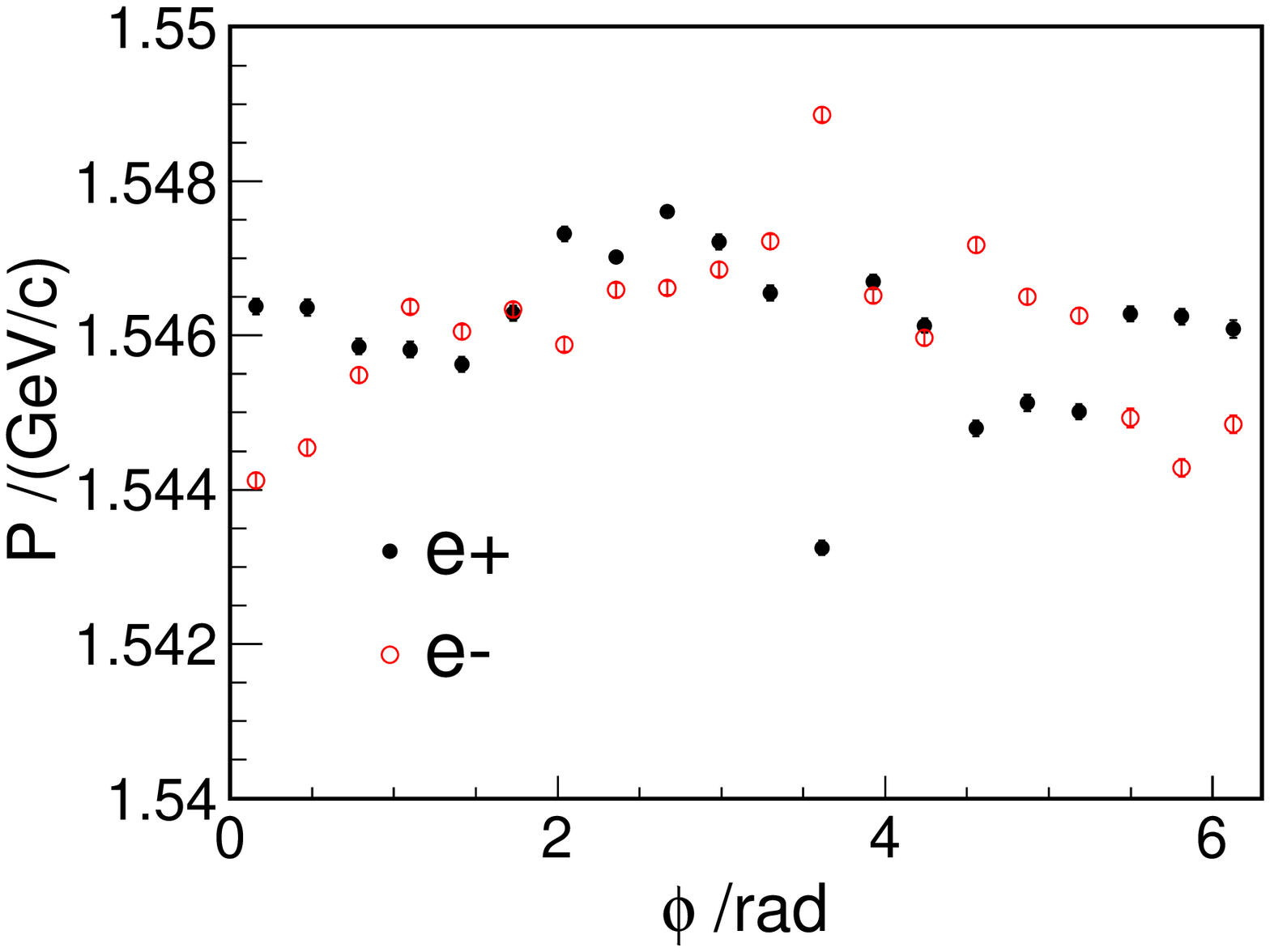}\put(-130,90){(a)}

  \includegraphics[width=6cm,height=4cm]{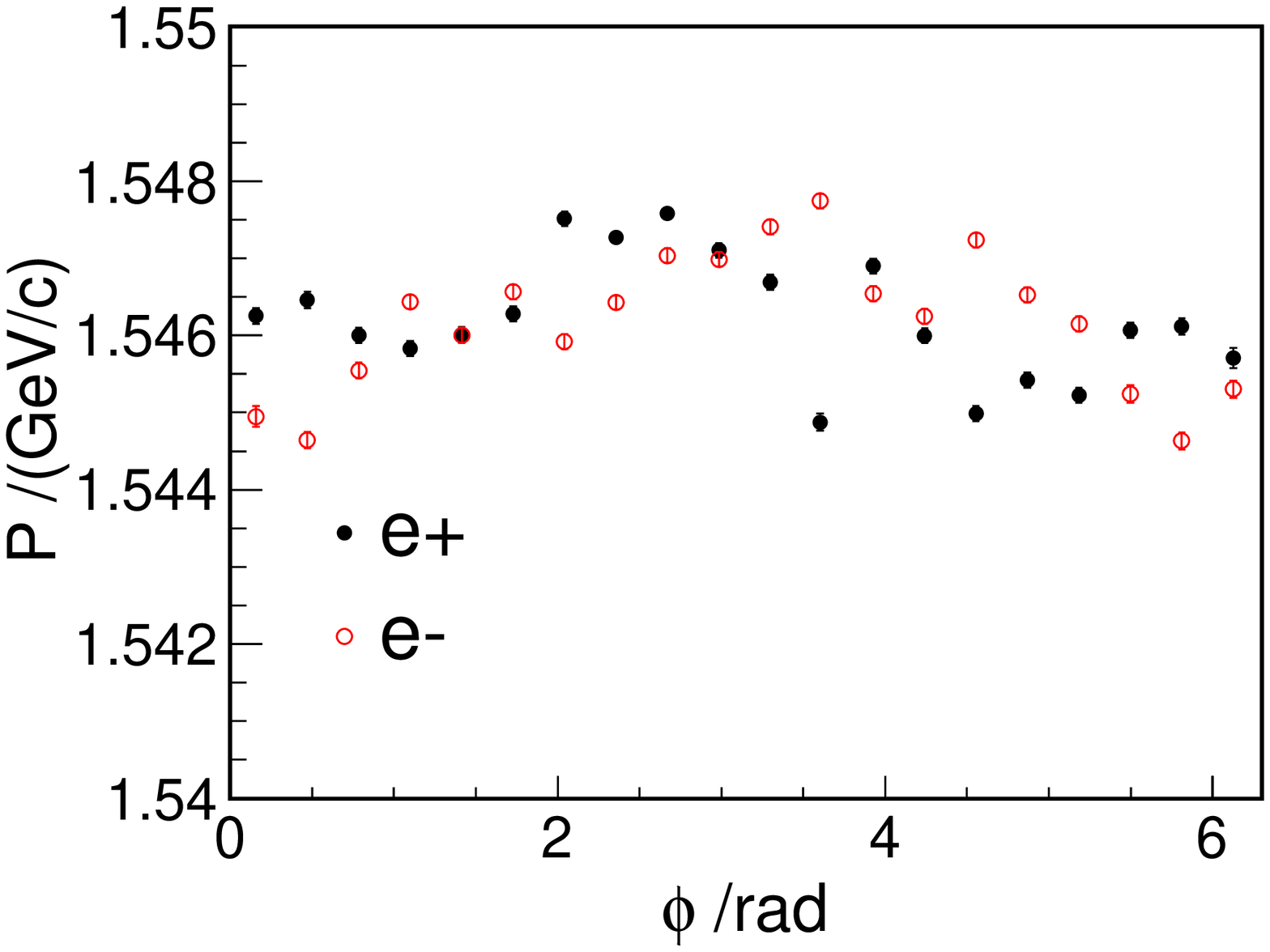}\put(-130,90){(b)}
  \figcaption{\label{resi_mom} The momentum as a function of $\phi$ for electron and positron ignoring (a) and considering (b) dead channels.}
\end{center}

In order to minimize this bias, we calibrated the affected cells in layer 42 separately. The variation of the X-T relation obtained from data is similar to the simulation result. The updated residual distributions are presented in Fig.~\ref{resi_deadch} (b), where the offset disappears. The difference in the reconstructed momentum for electrons and positrons in the affected region ($\phi\simeq2.75$ rads) is also reduced, as shown in Fig.~\ref{resi_mom} (b). Similar results are obtained for Dimuon event samples. However, the reconstructed momentum is still not flat in the azimuth direction, which may be related to misalignment or some other reasons, a subject of continuing studies.

\section{Spatial resolution}

After the complete calibration procedure including the calibration of X-T relations, time offset, time walk effects and alignment was done, the spatial resolution was determined. Here, as an example, we report the result for the J/$\psi$ data set collected in 2012. The spatial resolution, obtained from fitting a double Gaussian to the residual distribution, is 123 $\mu$m for Bhabha tracks averaged over the entire cells in all layers, as shown in Fig.~\ref{spa_reso}. Figure \ref{mom_reso} shows the spatial resolution as a function of drift distance in layer 25. For tracks traversing the middle region between the sense and field wire, the spatial resolution is better than 65 $\mu$m, which is close to the intrinsic resolution of the chamber.

\begin{center}
  \includegraphics[width=6cm,height=4cm]{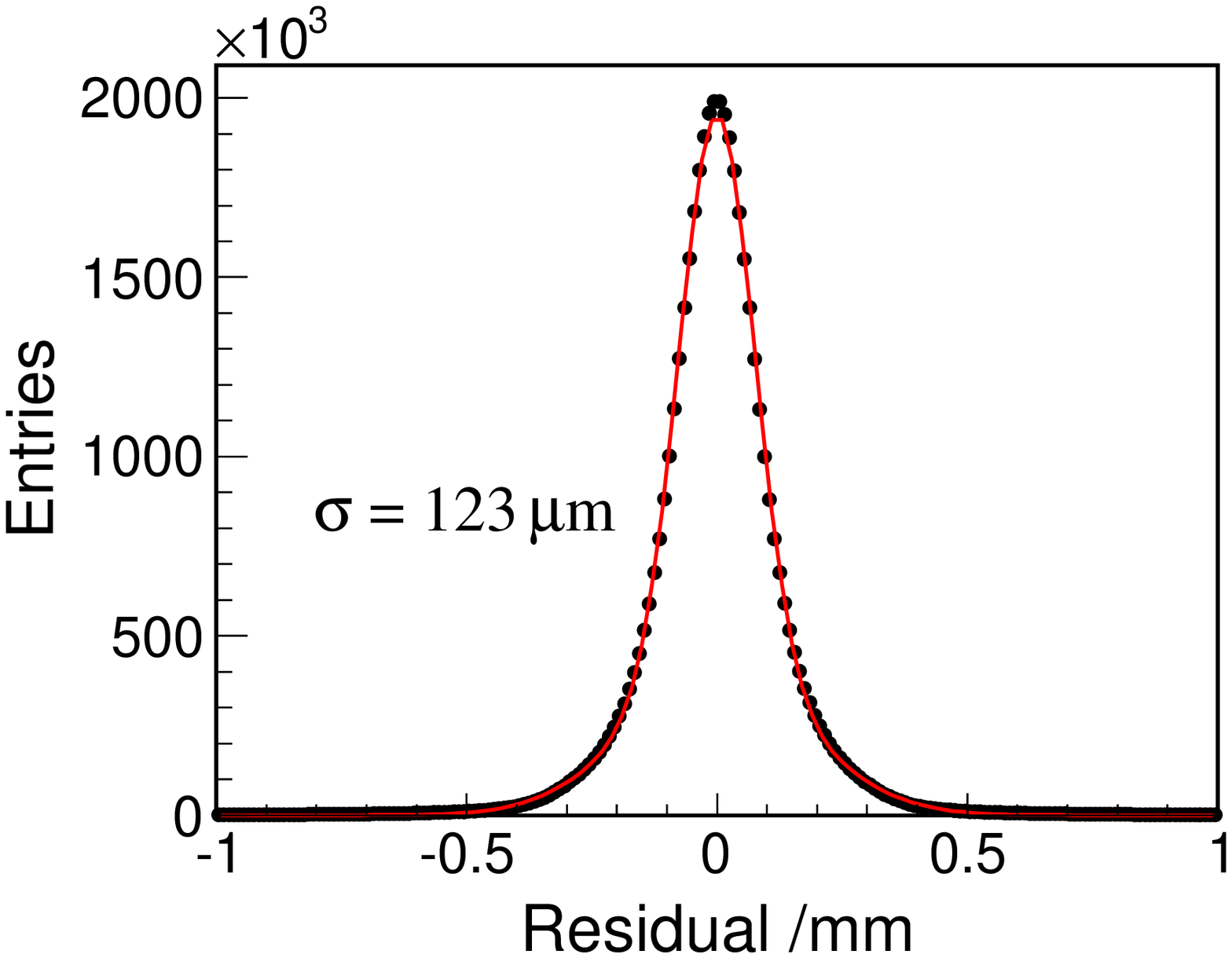}
\figcaption{\label{spa_reso} Spatial resolution for 1.548 GeV/c electrons from Bhabha events.}
\end{center}

\begin{center}
\includegraphics[width=6cm,height=4cm]{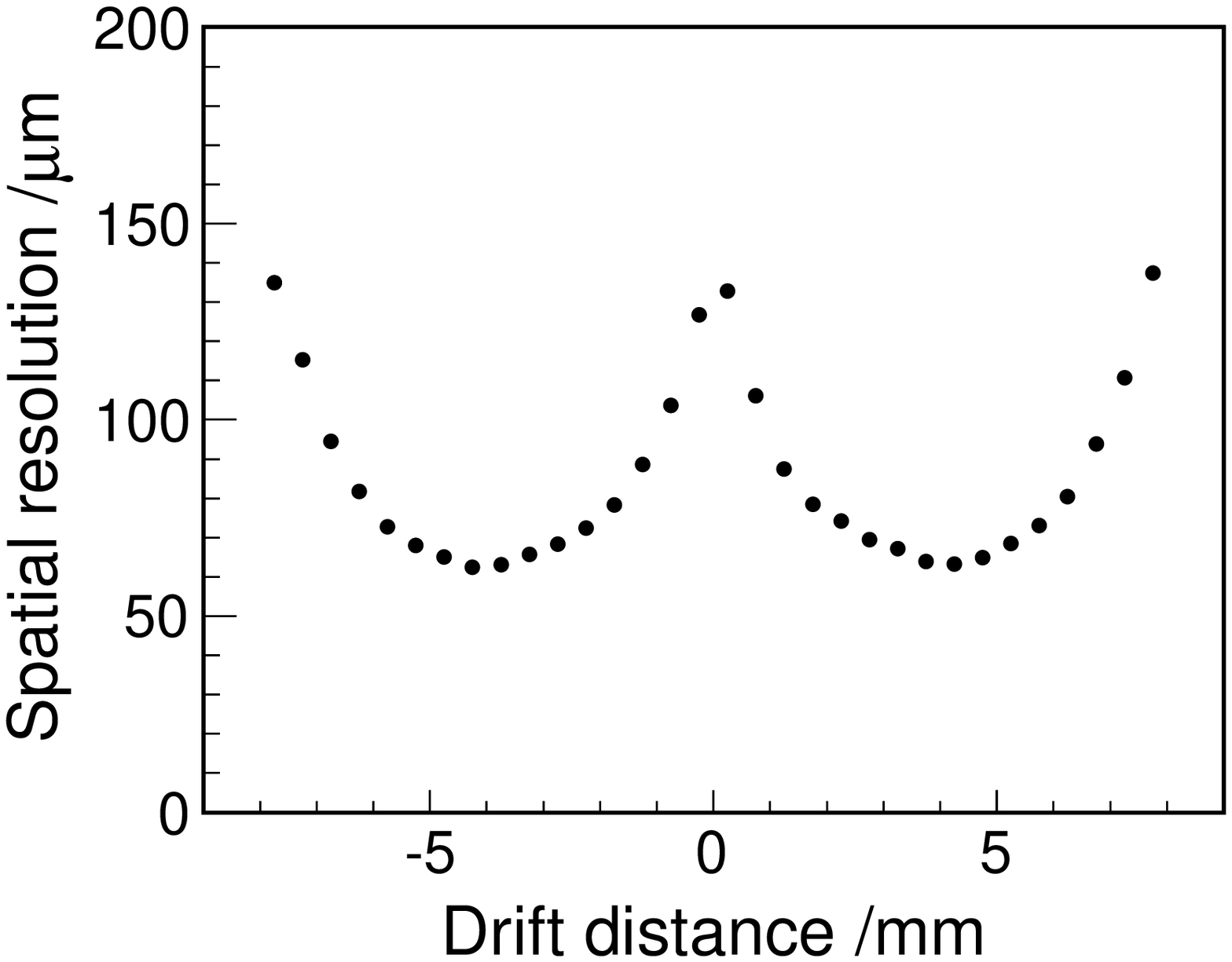}
\figcaption{\label{mom_reso} Spatial resolution as a function of drift distance in layer 25.}
\end{center}

\section{Summary}

The parameterization of X-T relation and its calibration procedure for BES\uppercase\expandafter{\romannumeral3} drift chamber is described in detail. In order to get accurate position information, efforts have been done from several aspects. First,
we use separate X-T relations for left and right and also for different entrance angles which are related to the track's momentum. We use low momentum tracks combined with Bhabha samples to calibrate X-T relations for tracks with large entrance angles. Next, to eliminate the impact of dead channels on X-T relation, we recalibrate the cells affected by adjacent dead channels. Finally, we determine the spatial resolution (123 $\mu$m) for Bhabha tracks (from J/$\psi$ data set collected in 2012) in the BES\uppercase\expandafter{\romannumeral3} drift chamber.

\end{multicols}

\vspace{-1mm}
\centerline{\rule{80mm}{0.1pt}}
\vspace{2mm}

\begin{multicols}{2}

\end{multicols}

\clearpage
\end{document}